\documentclass[runningheads]{llncs}
\usepackage[T1]{fontenc}
\usepackage{graphicx}
\usepackage{cite}
\usepackage{amsmath,amssymb,amsfonts}
\usepackage{algorithmic}
\usepackage{graphicx}
\usepackage{textcomp}
\usepackage{xcolor}
\usepackage{comment}
\usepackage{booktabs}
\usepackage{subfig}
\usepackage{array}
\usepackage{tabularx}
\newcolumntype{Y}{>{\centering\arraybackslash}X}
\begin{document}
\title{Serial-Parallel Dual-Path Architecture for Speaking Style Recognition}
\author{Guojian Li, Qijie Shao, Zhixian Zhao, Shuiyuan Wang, Zhonghua Fu, \and Lei Xie$^{*}$}
\authorrunning{Guojian Li et al.}
\institute{Audio, Speech and Language Processing Group (ASLP@NPU),\\
School of Computer Science, Northwestern Polytechnical University, Xi’an, China \\
\url{http://www.npu-aslp.org/}\\
\email{aslp\_lgj@mail.nwpu.edu.cn, lxie@nwpu.edu.cn}}
\maketitle              % typeset the header of the contribution
\begin{abstract}
Speaking Style Recognition (SSR) identifies a speaker's speaking style characteristics from speech. Existing style recognition approaches primarily rely on linguistic information, with limited integration of acoustic information, which restricts recognition accuracy improvements. The fusion of acoustic and linguistic modalities offers significant potential to enhance recognition performance. In this paper, we propose a novel serial-parallel dual-path architecture for SSR that leverages acoustic-linguistic bimodal information. The serial path follows the ASR+STYLE serial paradigm, reflecting a sequential temporal dependency, while the parallel path integrates our designed Acoustic-Linguistic Similarity Module (ALSM) to facilitate cross-modal interaction with temporal simultaneity. Compared to the existing SSR baseline---the OSUM model, our approach reduces parameter size by 88.4\% and achieves a 30.3\% improvement in SSR accuracy for eight styles on the test set.

\keywords{Speaking Style Recognition, Serial-Parallel, Cross-Modal, Acoustic-Linguistic Similarity, ASR + STYLE.}
\end{abstract}

\begingroup
\renewcommand\thefootnote{$^{*}$}
\footnotetext{Corresponding author.}
\endgroup

\section{Introduction}
Human speech not only conveys linguistic information but also contains rich paralinguistic cues, such as style, accent, and emotion~\cite{lin2024paralinguistics}~\cite{wang2018style}. Speaking Style Recognition (SSR) aims to extract speaking style features from speech signals, serving as a crucial technique for distinguishing different modes of speaking expression. %By capturing specific patterns in speech, SSR supports applications in speech understanding, personalized dialogue, and content recommendation~\cite{chu2024qwen2}~\cite{yu2024salmonn}~\cite{chen2025minmo}~\cite{gao2021advances}, significantly enhancing user experience. 
The potential value of SSR is gradually emerging across multiple domains~\cite{chu2024qwen2,yu2024salmonn,chen2025minmo,gao2021advances}. In speech understanding, SSR enhances the system's contextual sensitivity, thereby optimizing user experience. In intelligent dialogue systems, SSR facilitates personalized interactions, making the system more human-like. In content recommendation, SSR tailors recommendations to users' preferred speaking styles, improving relevance and recommendation effectiveness. As a highly promising avenue in speech processing and artificial intelligence, SSR warrants further attention and exploration.

Current research on style recognition primarily focuses on the text domain~\cite{tang2022naughtyformer,christ2024towards,fang2024single,kamal2019self}, with mainstream methods relying solely on linguistic modalities, such as lexicon and semantics. For example, in humour style recognition, Kenneth et al.~\cite{kenneth2024two} proposed a two-model cascading architecture, integrating DistilBERT and statistical features, which achieved remarkable performance in distinguishing between affiliative and aggressive humour styles. In multi-authored style shift detection, Zamir et al.~\cite{zamir2024stylometry} innovatively utilized non-semantic features such as punctuation distribution and special character sequences, combined with a BiLSTM-CRF hybrid model, to achieve high performance in detecting style boundaries in multi-authored documents. In academic writing integrity assessment, Oliveira et al.~\cite{oliveira2025human} introduce an authorship verification framework that constructs individualized writing profiles to quantify AI assistance in student compositions. By adapting the Feature Vector Difference method, their framework captures stylometric nuances at both word and sentence levels, effectively distinguishing between human-authored and AI-generated texts.

Most existing style recognition approaches rely on linguistic information. While they have proven effective, their lack of integration with acoustic information limits their performance potential. OSUM~\cite{geng2025osum}, an open speech understanding model, pioneered the exploration of the speaking style recognition (SSR) task by adopting the ASR+STYLE serial paradigm, where the large language model (LLM) initially generates automatic speech recognition (ASR) transcriptions and subsequently integrates them with acoustic features to predict style labels. This attempt integrates acoustic and linguistic information to some extent, improving the performance of speaking style recognition. However, we argue that the ASR+STYLE serial paradigm alone is insufficient for fully integrating bimodal information, and there remains substantial potential for enhancing the performance of speaking style recognition.

In this paper, we leverage both acoustic and linguistic bimodal information to propose a novel serial-parallel dual-path architecture for the SSR task. Specifically, we construct the serial path by drawing on the ASR+STYLE serial paradigm~\cite{geng2025osum}. In this path, the LLM first generates the ASR transcriptions and then combines them with acoustic features to infer style recognition labels. This step-by-step process creates a clear temporal order and stage-wise dependence in the time dimension, which is why it is termed “serial”. Simultaneously, inspired by the concept of linguistic-acoustic similarity from~\cite{shao2022lasas}, we design an Acoustic-Linguistic Similarity Module (ALSM) as the parallel path. In contrast to the serial path, the parallel path %abandons temporal dependencies, 
emphasizes the simultaneous processing of acoustic and linguistic features by highlighting the temporal synchronization in cross-modal interaction, thus being termed “parallel”. Experimental results show that, compared to the existing SSR baseline---the OSUM model, our proposed approach reduces the parameter size by 88.4\% while achieving a 30.3\% improvement in SSR accuracy across eight styles on the test set, achieving efficient fusion of acoustic and linguistic bimodal information.

\begin{figure*}[t]
    \centering
    \includegraphics[width=\linewidth,keepaspectratio]{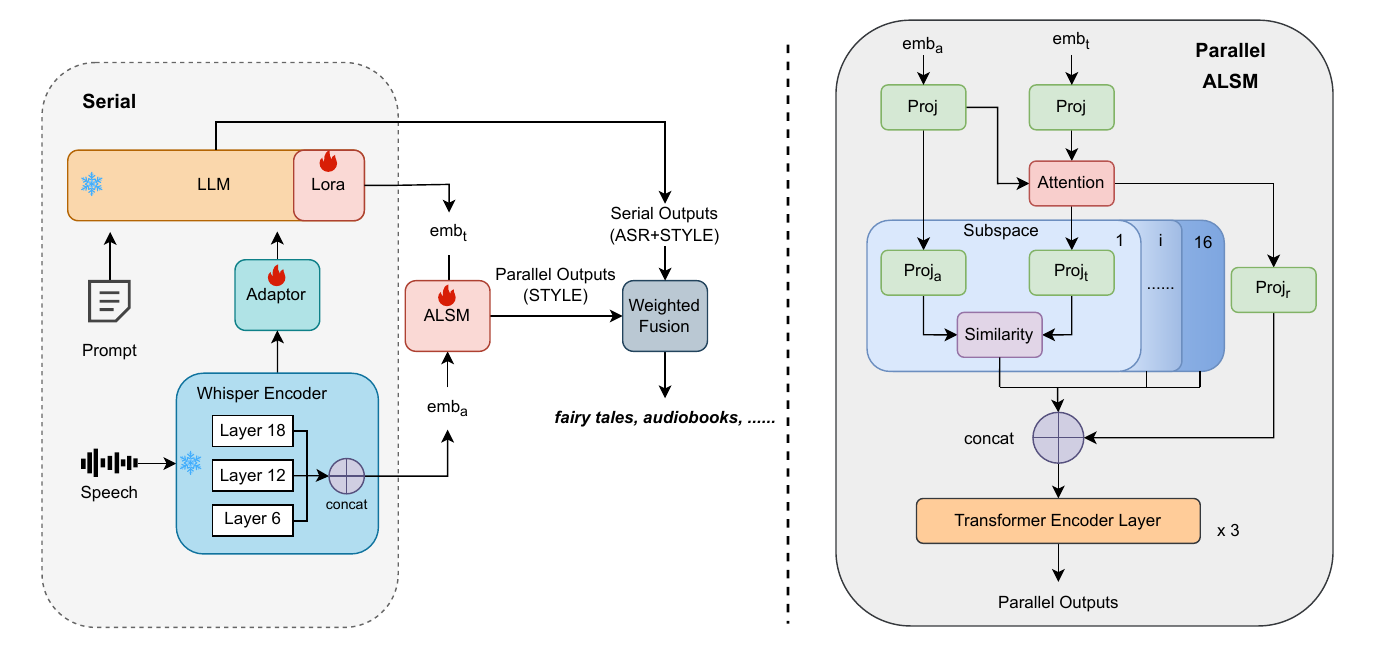}
    \caption{The overall architecture of the proposed model and ALSM.}
    \label{fig:overview}
\end{figure*} 

\section{METHODS}

\subsection{Overview}
As shown in Figure~\ref{fig:overview}, this paper proposes a serial-parallel dual-path architecture, aiming to enhance SSR performance by leveraging both acoustic and linguistic bimodal information. 

Specifically, the serial path adopts a three-component architecture, including an acoustic encoder, adaptor module, and LLM~\cite{geng2024unveiling}. It follows the ASR+STYLE serial paradigm introduced in the OSUM~\cite{geng2025osum}: the LLM first produces ASR transcriptions, which are subsequently integrated with acoustic features to infer style labels. This sequential process establishes a distinct temporal order and stage-wise dependency along the time axis. In contrast, the parallel path utilizes the Acoustic-Linguistic Similarity Module (ALSM), which simultaneously inputs acoustic and linguistic features. By aligning them along the time dimension through an attention mechanism, ALSM computes cross-modal similarity to infer style labels. The parallel path emphasizes the synchronized processing of acoustic and linguistic features, reflecting a temporally synchronized cross-modal interaction. Next, we perform weighted fusion of the outputs from the serial and parallel paths, combining the advantages of both paths and fully leveraging the acoustic-linguistic bimodal information to obtain the final style recognition result. The details of each part are as follows.

\subsection{Serial Path}
For the serial path, we employ the ASR+STYLE serial paradigm~\cite{geng2025osum}, where the LLM generates ASR transcriptions and combines them with acoustic features to predict style labels sequentially, establishing a sequential temporal dependency. To operationalize the ASR+STYLE serial paradigm, we employ natural language prompts to instruct the LLM. ChatGPT~\cite{achiam2023gpt} is utilized to generate five candidate prompts, with one randomly selected during training. The input to the serial path consists of speech signals and natural language prompts, while the output corresponds to the results generated by the ASR+STYLE serial paradigm. An example of the prompts and the ASR+STYLE prediction results is illustrated in Table~\ref{tab:prompt}.

\begin{table}[t]
  \centering
   \caption{An illustration of the prompts and the corresponding ASR+STYLE prediction results.}
  \begin{tabularx}{\columnwidth}{@{} Y | Y @{}}
    \toprule
    \textbf{Prompt} 
      & \textbf{ASR+STYLE prediction result} \\
    \midrule
    Please transcribe the audio into text and append a 
    \textless{}style\textgreater{} label at the end of the transcription. 
    The available style labels include: 
    \textless{}news and science reporting\textgreater{}, 
    \textless{}horror stories\textgreater{}, 
    \textless{}fairy tales\textgreater{}, 
    \textless{}customer service\textgreater{}, 
    \textless{}poetry and prose\textgreater{}, 
    \textless{}audiobooks\textgreater{}, 
    \textless{}spontaneous conversation\textgreater{}, 
    and \textless{}others\textgreater{}. 
      & The Little Thumbelina did not like her mole neighbor at all. \textless{}fairy tales\textgreater{} \\
    \bottomrule
  \end{tabularx}
  \label{tab:prompt}
\end{table}

The serial path consists of three core components: an acoustic encoder, an adaptor module, and an LLM. During training, the encoder remains frozen, the adaptor module parameters are fully updated, and the LLM is fine-tuned using LoRA~\cite{hu2022lora}. We employ the Whisper-Medium model~\cite{radford2023robust} as the acoustic encoder, featuring two layers of 1D convolution with 2 times downsampling, followed by 24 Transformer layers, totaling approximately 300 million parameters. The adaptor module, with around 50 million parameters, adopts a hybrid architecture~\cite{xu2025adaptor} combining three layers of 1D convolution and four Transformer layers. Unlike OSUM, which utilizes the large-scale Qwen2-7B-Instruct LLM~\cite{an2024qwen2}  for multi-speech understanding tasks, our approach focuses on the SSR task, opting for the smaller Qwen2.5-0.5B-Instruct LLM~\cite{an2024qwen2.5} to enhance training and inference efficiency.

\subsection{Parallel Path}

We design an Acoustic-Linguistic Similarity Module (ALSM) as the parallel path, consisting of four critical stages: Bimodal feature projection aligns the acoustic and linguistic features to the same feature dimension. Attention-guided alignment ensures temporal synchronization between the two modalities. Multi-space decoupled cross-modal similarity measurement facilitates cross-modal interaction with temporal simultaneity. Transformer classification produces the final style prediction labels. The detailed architecture of the ALSM is illustrated in Figure~\ref{fig:overview}.

\subsubsection{Bimodal Feature Projection}

We leverage the Whisper-medium audio encoder~\cite{radford2023robust} to extract acoustic features. To bridge the gap between low-level acoustic details and high-level linguistic semantics~\cite{yuan2023whisper-at}, we select the hidden states from layers 6, 12, and 18 and concatenate them, generating an acoustic embedding $emb_a$ with  a dimension of 3072 ($1024 \times 3$) to integrate information from different layers. Subsequently, it is passed through a linear projection layer $\text{Proj}$ with a GELU activation function, reducing the dimension to 256. Layer Normalization is then applied to eliminate discrepancies in feature distributions, yielding the final acoustic feature $h_a$:
\begin{equation}
h_a = \text{LayerNorm}(\text{GELU}(\text{Proj}(emb_a))) \in \mathbb{R}^{B \times T \times D}
\label{eq:acoustic_feature},
\end{equation}
where $B$ denotes batch size, $T$ the number of audio frames, and $D=256$.

For linguistic feature processing, we use the last layer hidden state of the LLM output as the linguistic embedding  $emb_t$, with a dimension of 896. The embedding is then processed through steps similar to those applied to the acoustic embedding, resulting in a 256-dimensional linguistic feature $h_t$, aligned in feature dimension with the acoustic feature $h_a$, as detailed below:
\begin{equation}
h_t = \text{LayerNorm}(\text{GELU}(\text{Proj}(emb_t))) \in \mathbb{R}^{B \times S \times D},
\label{eq:text_feature}
\end{equation}
where $B$ denotes batch size, $S$ the number of text tokens, and $D=256$.

\subsubsection{Attention-guided Alignment}
To achieve temporal simultaneity for subsequent cross-modal interactions, we utilize an attention mechanism to align the text token sequence with the audio frame sequence along the temporal dimension. Specifically, attention weights are computed to quantify the relevance between each audio frame and text token. These weights are then applied to perform a weighted summation of the linguistic feature $h_t$, yielding a linguistic feature $h_{t\text{-}al}$ temporally aligned with the acoustic feature $h_a$, as formulated in the following equation:
\begin{equation}
h_{t\text{-}al} = \text{Softmax}\left( \frac{h_a \cdot h_t^\top}{\sqrt{d}} \right) \cdot h_t \quad \in \mathbb{R}^{B \times T \times D}.
\label{eq:attention_alignment}
\end{equation}

\subsubsection{Multi-Space Decoupled Cross-Modal Similarity Measurement}
To decouple style-specific acoustic-linguistic interaction patterns, we project the aligned acoustic ($\mathbf{h}_a$) and linguistic ($\mathbf{h}_{t\text{-}al}$) features into $N$ distinct latent subspaces ($N = 16$, empirically chosen based on~\cite{shao2022lasas}) via linear projections. For each subspace $i$ $\in \{1,\ldots,N\}$:
\begin{equation}
\begin{aligned}
\mathbf{h}_a^{(i)} &= \text{Proj}_{a}^{(i)}(\mathbf{h}_a) \in \mathbb{R}^{B \times T \times D} \\
\mathbf{h}_{t\text{-}al}^{(i)} &= \text{Proj}_{t}^{(i)}(\mathbf{h}_{t\text{-}al}) \in \mathbb{R}^{B \times T \times D},
\end{aligned}
\label{eq:multi_space}
\end{equation}
where $\text{Proj}_{a}^{(i)}$ and $\text{Proj}_{t}^{(i)}$ denote trainable linear projection layers for subspace $i$, preserving the feature dimension $D = 256$. The dimension preservation ($D \rightarrow D$) ensures structural consistency of cross-modal interactions and prevents information loss during subspace projection process. 

Then, we compute the cosine similarity $s_i$ between the projected acoustic features $h_a^{(i)}$ and linguistic features $h_{t-al}^{(i)}$ within each decoupled subspace to measure the local cross-modal alignment strength ~\cite{langari2005hybrid}\cite{guzhov2022audioclip}. The cross-modal correlation in subspace $i$ is formulated as:
\begin{equation}
s_i = \frac{h_a^{(i)} \cdot h_{t\text{-}al}^{(i)}}{\|h_a^{(i)}\|_2 \|h_{t\text{-}al}^{(i)}\|_2} \in \mathbb{R}^{B \times T}.
\label{eq:cos_sim}
\end{equation}

Additionally, we introduce a semantic preservation branch~\cite{shao2022lasas}, which generates the reduced-dimensional linguistic reference feature $h_{t-al}^{ref}$ through a linear projection layer $\text{Proj}_r$ and the activation function $\text{Tanh}$, effectively retaining the original linguistic information of the text: 
\begin{equation}
h_{t-al}^{ref} = \text{Tanh}(\text{Proj}_r(h_{t-al})) \in \mathbb{R}^{B \times T \times 128}.
\label{eq:semantic_preserve}
\end{equation}

Subsequently, we concatenate the acoustic-linguistic similarities ${\{s_i\}}_{i=1}^N$ with the reduced-dimensional linguistic reference feature $h_{t-al}^{ref}$ to construct the cross-modal acoustic-linguistic representation $h_{\text{cm}}$:
\begin{equation}  
h_{\text{cm}} = \left[ s_1 \mid \cdots \mid s_N \mid h_{t\text{-}al}^{\text{ref}} \right] \in \mathbb{R}^{B \times T \times (N+128)}. 
\label{eq:bimodal_rep}  
\end{equation}

\subsubsection{Transformer Classification}

Finally, we leverage a three-layer Transformer encoder to process the cross-modal acoustic-linguistic representation $h_{\text{cm}}$, %obtained earlier, 
capturing frame-wise temporal dependencies and extracting essential features from both acoustic and linguistic modalities. The encoded outputs are processed through global average pooling and normalized via the LogSoftmax function, producing a probability distribution over eight style categories. The final style prediction label is then determined by selecting the category with the highest probability.

\begin{figure}[h]
    \centering
    \includegraphics[width=1\columnwidth]{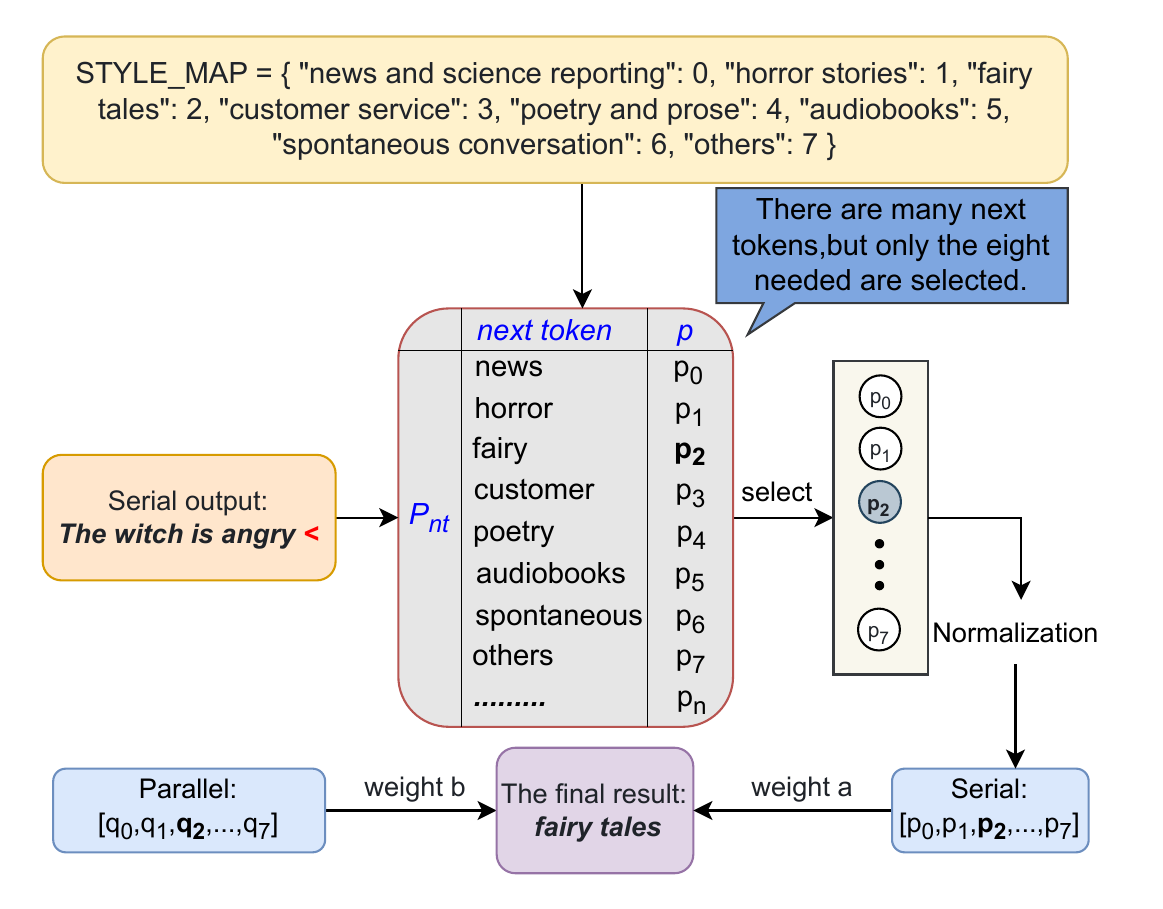}
    \caption{The inference process of our proposed model.}
    \label{fig:infer}
\end{figure}

\subsection{Combination of Serial and Parallel Paths}

\subsubsection{Training}

Our proposed model employs a serial-parallel dual-path optimization strategy during training, with both paths relying on cross-entropy loss for optimization. The serial path generates ASR+STYLE outputs by first producing ASR transcriptions, followed by style recognition results, from which the loss term $\mathcal{L}_{\text{CE}}^{\text{serial}}$ is computed. The parallel path directly predicts style recognition results, yielding the $\mathcal{L}_{\text{CE}}^{\text{parallel}}$ term. The overall loss function is weighted by coefficients $\alpha$ and $\beta$, and is formulated as:
\begin{equation}
\mathcal{L}_{\text{total}} = \alpha \cdot \mathcal{L}_{\text{CE}}^{\text{serial}} + \beta \cdot \mathcal{L}_{\text{CE}}^{\text{parallel}}.
\label{eq:total_loss}   
\end{equation}

\subsubsection{Inference}

During inference, our proposed model performs weighted fusion of the outputs from the serial and parallel paths, effectively leveraging the strengths of both paths, thereby significantly enhancing the accuracy of speech style recognition. The inference process of our proposed model is illustrated in Figure~\ref{fig:infer}.

The LLM in the serial path functions as an autoregressive model, generating each token sequentially by conditioning on previously produced tokens until a termination token (e.g., eos), signals the end of the sequence. In contrast, the parallel path produces non-autoregressive outputs, making direct weighted fusion impractical. To address this, we implement specialized processing on the autoregressive output from the serial path.

We designate the symbol “\textless” as a special termination token to truncate sequence generation and identify the starting position of style labels. Through fine-tuning on our style data, we observed that after generating the termination token “\textless”, the LLM in the serial path exclusively produces one of the eight predefined style labels in the style mapping dictionary STYLE\_MAP (as illustrated in Figure~\ref{fig:infer}). Furthermore, while each style label consists of one or more tokens, the first token of every style label has been ensured to be unique and distinct across all eight style labels during both training and inference. Leveraging the above properties, we calculate probabilities by extracting only the probability of the first token for each style label, rather than the entire label, thereby simplifying the computational process and significantly improving inference efficiency.

Specifically, during the autoregressive generation process, once the LLM in the serial path generates the special termination token “\textless”, the probability distribution of the next token $P_{nt}$ is immediately computed. From this distribution, the probabilities associated with the first tokens of the eight style labels defined in the STYLE\_MAP are selected (e.g., $p_2$ represents the probability of the first token for the “fairy tales” style label, as shown in Figure~\ref{fig:infer}). These probabilities are then normalized to form a probability distribution $[p_0, p_1, p_2, ..., p_7]$ representing the eight style categories.

Different from the method employed in the serial path, the parallel path employs a three-layer Transformer encoder to process the fused acoustic and linguistic bimodal features, directly %(non-autoregressively) 
outputting a probability distribution $[q_0, q_1, q_2, ..., q_7]$ corresponding to the eight style categories.

Finally, the style probability distributions generated by the serial and parallel paths are linearly combined with weights $a$ and $b$, respectively. The category with the highest combined probability is selected as the final style prediction result.

\section{EXPERIMENTAL SETUP}

\subsection{Data Preparation} %数据构建

Our speech style recognition training set consists of two components. First, we include 315 hours of high-quality internal data with manually annotated style labels, spanning five categories: news and science reporting, poetry and prose, audiobooks, customer service, and fairy tales.
Second, to address the scarcity of labeled style data, we filter and annotate additional unlabeled internal data. Using the openSMILE~\cite{eyben2013recent} and Librosa~\cite{mcfee2015librosa} toolkits, we extract five acoustic features: speaking rate, energy, energy standard deviation, pitch mean, and pitch standard deviation. Each feature is divided into low, medium, and high intervals through statistical binning~\cite{schuller2012automatic}, based on the global mean and standard deviation. Data where all five features fall into the high interval are filtered as high-expressivity style candidates. Then, we utilize two open-source LLMs, Qwen2.5-14B-Instruct~\cite{an2024qwen2.5} and GLM-4-9B-Chat~\cite{glm2024chatglm}, to annotate the highly expressive data obtained from the previous feature filtering step. Qwen2.5-14B-Instruct represents a widely adopted model in the Qwen2.5 series, while GLM-4-9B-Chat is an open-source iteration of Zhipu AI’s GLM-4 series. Due to the absence of open-source tools for audio style annotation, these LLMs are employed with carefully designed prompts to annotate the textual transcripts of style data. To ensure annotation quality, we consider only the intersection of labels produced by both models, resulting in 1,950 hours of high-expressivity labeled data. By combining both components, our final dataset encompasses 2,265 hours of speech data, categorized into eight styles aligned with the OSUM taxonomy~\cite{geng2025osum}: news and science reporting, horror stories, fairy tales, customer service, poetry and prose, audiobooks, spontaneous conversation, and others.

Our speech style recognition test set comprises 3,000 samples, encompassing eight style categories consistent with those in the training set. The data sources for the test set are twofold: one part consists of internally curated high-quality style-labeled data, while the other part is derived from unlabeled style data processed through joint annotation by the aforementioned two LLMs, followed by manual evaluation. The manual evaluation process is as follows: two independent reviewers assess the LLM-generated style labels based on the original audios and corresponding transcripts, assigning confidence scores ranging from 1 to 10. The average of the two reviewers’ scores is calculated, and only samples with an average score greater than 5 are retained~\cite{zhou2025calibrating}. This process is designed to enhance the quality of the test set.

\subsection{Implement Details}
During training, the proposed model is optimized using the
AdamW optimizer with an initial learning rate of \(5.0 \times 10^{-5}\) and a batch size of 16. We freeze the Whisper-Medium encoder (initialized from the Whisper-Medium encoder in open-source OSUM~\cite{geng2025osum}) 
while training the adaptor module, ALSM and LLM. The LLM is fine-tuned with LoRA~\cite{hu2022lora}, with the LoRA rank set to 8, the scaling factor set to 32, and the dropout rate for LoRA matrices set to 0.1. The loss function (shown in Eq.~\eqref{eq:total_loss}) is defined with weights $\alpha=1.0$ for $\mathcal{L}_{\text{CE}}^{\text{serial}}$ and $\beta=0.5$ for $\mathcal{L}_{\text{CE}}^{\text{parallel}}$. Training spans 10 epochs on 4 NVIDIA RTX 4090 GPUs using the WeNet toolkit~\cite{yao2021wenet}.

At inference time, the final style prediction result is obtained by fusing the outputs of the serial and parallel paths with weights $a=0.3$ and $b=0.7$, respectively.

\subsection{Comparison Systems}
To assess the performance of our proposed approach, we implement the following systems. 
\begin{itemize}
\item \textbf{GLM-4-9B-Chat}: GLM-4-9B-Chat~\cite{glm2024chatglm} is a prominent text-based LLM in the ChatGLM series, built on a Transformer decoder architecture. It excels in generating long-form text and interpreting complex instructions.
\item \textbf{Qwen2.5-0.5B-Instruct}: Qwen2.5-0.5B-Instruct~\cite{an2024qwen2.5} is a lightweight text-based LLM in the Qwen2.5 series, leveraging a standard Transformer decoder architecture. It is optimized for instruction following in resource-constrained environments.
\item \textbf{Whisper-Medium Encoder + FC}: The Whisper-Medium encoder~\cite{radford2023robust} contains two layers of 1D convolution %2 one-dimensional convolutional layers (2 × stride) 
and 24 transformer layers. A single fully-connected layer serves as the classifier. This combination represents a widely adopted approach in audio classification tasks. The total number of parameters is approximately 0.3 billion.
\item \textbf{OSUM}: The OSUM model~\cite{geng2025osum} is an open-source multi-task speech understanding model that pioneers the exploration of the SSR task. It integrates three components: a Whisper-Medium encoder, an adaptor module, and the Qwen2-7B-Instruct LLM. The total parameter count reaches approximately 7.35 billion.
\end{itemize}

\begin{table}[htbp]
\caption{Performance comparison of different approaches. }%“Fine-tuned” indicates whether the model is fine-tuned on our style data. “Text-only” denotes the use of only textual transcripts from the style data. "Audio-only" signifies the use of only the style audios. "Audio+Text" represents the combined use of both style audios and their corresponding textual transcripts.
\label{table:results}
\centering
\resizebox{\textwidth}{!}{
\setlength{\tabcolsep}{6pt}
\renewcommand{\arraystretch}{1.3} % 行距
\begin{tabular}{clcccc}
\toprule
\textbf{Exp. ID} & \textbf{Model Name} & \textbf{Fine-tuned(Yes/No)} & \textbf{Params} & \textbf{Modality} & \textbf{Accuracy (\%) $\uparrow$ } \\
\midrule
E1 & GLM4-9B-Chat & No & 9B & Text-only & 53.97 \\
E2 & Qwen2.5-0.5B-Instruct & No & 0.5B & Text-only & 
37.90 \\
E3 & Qwen2.5-0.5B-Instruct & Yes & 0.5B & Text-only & 
66.87 \\
E4 & Whisper-Medium Encoder + FC & Yes & 0.3B  & Audio-only & 66.60 \\
E5 & OSUM  & Yes & 7.35B & Audio+Text & 67.34 \\
E6 & Proposed Model & Yes & 0.855B & Audio+Text & \textbf{87.73} \\
\bottomrule
\end{tabular}
}
\end{table}

\section{EXPERIMENTAL RESULTS}
\subsection{Comparison of Different Approaches}

As shown in Table~\ref{table:results}, we compare the performance of different approaches on the test set, using accuracy as the evaluation metric. E1 and E2 utilize two open-source text-based LLMs to directly process textual transcripts from the test set for style recognition,  without any fine-tuning. E3 fine-tunes Qwen2.5-0.5B-Instruct exclusively on the textual transcripts from our style dataset, where the input is the text and the output is the corresponding style label. E4 fine-tunes the Whisper-Medium Encoder with a fully connected layer using only the audio samples from our style dataset, with audio as the input and only the style label as the output. E5 evaluates the SSR performance of the OSUM model, which fine-tunes on both audio samples and their associated textual transcripts from our style dataset. The OSUM model takes audio as input, generates ASR transcriptions, and subsequently predicts the style labels. E6 assesses the performance of our proposed approach.

Compared to the proposed approach E6, E1 and E2 show a significant gap in speaking style recognition accuracy. This indicates that while open-source text-based LLMs exhibit some capability in style recognition, their performance remains notably limited. 

A direct comparison of E3 with E5 and E6 demonstrates that, although fine-tuning on text style data significantly enhances the performance of pure text-based models in style recognition, their effectiveness still falls short of both the OSUM model and the proposed approach. This underscores that relying solely on linguistic modality is insufficient for achieving efficient speaking style recognition, and integrating acoustic modality is essential for superior performance. 

Furthermore, comparative analysis of E4 against both E5 and E6 reveals that the Whisper-Medium encoder, relying solely on the acoustic modality, performs less effectively than both the OSUM model and the proposed approach in the SSR task. These findings highlight that relying exclusively on the acoustic modality is inadequate for achieving satisfactory style recognition performance, and incorporating linguistic modality is a critical pathway for performance enhancement.

\begin{figure}[t]
    \centering
    \includegraphics[width=\columnwidth]{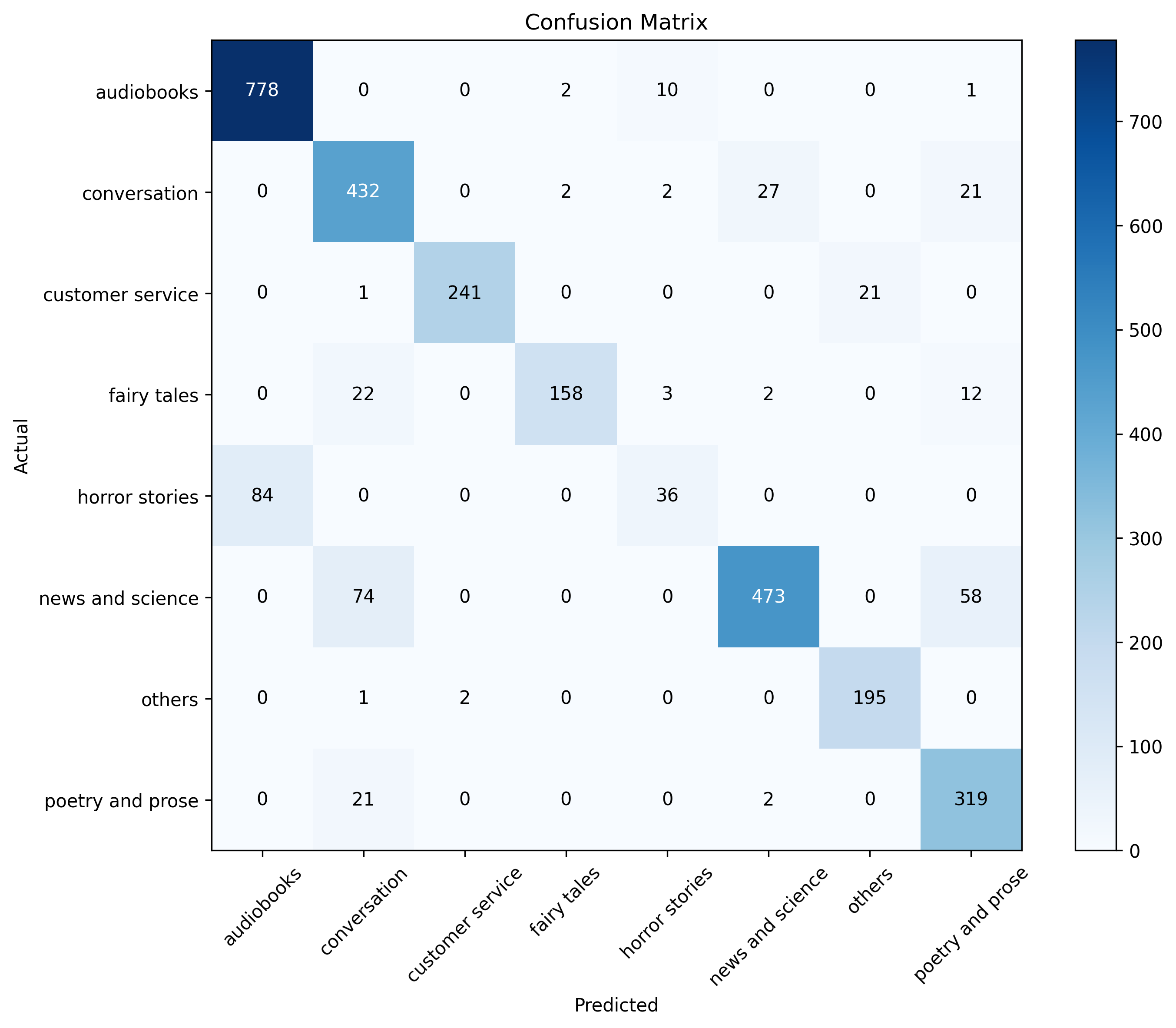}
    \caption{The confusion matrix for the eight styles of our proposed approach.}
    \label{fig:result}
\end{figure}

OSUM explores the integration of acoustic and linguistic modalities to some extent, achieving improved performance compared to single-modality approaches E3 and E4. However, its performance still leaves room for enhancement. Compared to OSUM, the proposed approach reduces the parameter count by 88.4\% while improving performance by 30.3\%. This demonstrates the effectiveness of the proposed serial-parallel dual-path architecture, which leverages acoustic-linguistic bimodal information for the SSR task. The detailed SSR results of the proposed approach are presented in Figure~\ref{fig:result} as a confusion matrix.

\subsection{Ablation Study}

\begin{table}[htbp]
\caption{Performance comparison of ablation components.}
\label{tab:ablation}
\centering
\scalebox{1.0}{ % 根据需要调整表格的缩放大小
\begin{tabular}{ccc}
\toprule
\textbf{Exp. ID} & \textbf{Model} & \textbf{Accuracy (\%) $\uparrow$ } \\
\midrule
E6 & Proposed Model & \textbf{87.73} \\
E7 & Serial: ASR+STYLE & 73.83 \\
E8 & Direct: Only STYLE & 67.99 \\
E9 & Parallel: ALSM & 84.17 \\
\bottomrule
\end{tabular}
}
\end{table}

We conducted ablation experiments on the proposed serial-parallel dual-path architecture, with the results summarized in Table~\ref{tab:ablation}, using accuracy as the evaluation metric. E6 represents our complete proposed architecture, which generates the final style recognition results through a weighted fusion of the serial and parallel paths. E7 evaluates the performance of the serial path in our proposed architecture, utilizing the ASR+STYLE serial paradigm. E8 employs the same three-component architecture as our serial path but does not adopt the ASR+STYLE paradigm, instead directly outputting style labels. E9 exclusively tests the performance of the parallel path---ALSM.

Experimental results show that E7 outperforms E8 in style recognition accuracy by 8.6\%, demonstrating the effectiveness of the ASR+STYLE serial paradigm. 
%where ASR transcriptions are first generated, followed by the output of style labels. 
Comparing E9 to E7, the parallel path outperforms the serial path, achieving a 14\% improvement, thereby validating the significant effectiveness of the designed ALSM as the parallel path in the SSR task. Furthermore, E6 achieves performance improvements of 18.8\% and 4.2\% compared to the standalone serial and parallel paths (E7, E9), respectively, highlighting the effectiveness and superiority of the proposed serial-parallel dual-path architecture.

\section{CONCLUSION}

In this paper, we propose a novel serial-parallel dual-path architecture for SSR that harnesses acoustic-linguistic bimodal information. The serial path follows the ASR+STYLE serial paradigm that reflects a sequential temporal dependency, while the parallel path integrates the Acoustic-Linguistic Similarity Module (ALSM) to facilitate synchronized cross-modal interactions. Experimental results demonstrate that our proposed architecture effectively improves speaking style recognition accuracy with a relatively small number of parameters. In future work, we aim to further enhance this serial-parallel dual-path bimodal architecture and extend its application to other speech understanding tasks such as emotion recognition and sound event detection.
\bibliographystyle{splncs04}
\bibliography{ref.bib}
\begin{comment}

\end{comment}
\end{document}